\documentclass[11pt,twoside]{article}
\usepackage{asp2010}
\usepackage{graphicx}
\newcommand*{\vcenteredhbox}[1]{\begingroup
\setbox0=\hbox{#1}\parbox{\wd0}{\box0}\endgroup}
\def\ltsim{\mathrel{\hbox{\rlap{\hbox{\lower4pt\hbox{$\sim$}}}\hbox{$<$}}}}

\resetcounters

\begin{document}

\title{Magnetism, rotation and large-scale wind variability of O-type stars}
\author{Gregg A. Wade
\affil{Dept. of Physics, Royal Military College of Canada, PO Box 17000, Station Forces, Kingston, ON, Canada, K7K 7B4}}

\begin{abstract}
The common - arguably ubiquitous - large-scale variability of optical and UV lines profiles of hot, massive stars is widely interpreted as the direct consequence of structured, variable winds. Many of the variability phenomena are observed to recur on timescales compatible with stellar rotation, suggesting a picture in which perturbations at the base of the wind - carried into view by stellar rotation - produce large-scale outward-propagating density structures. Magnetic fields have been repeatedly proposed to be at the root of these phenomena, although evidence supporting this view remains tenuous. In this review I discuss the evidence for large-scale structures in the winds of O-type stars, the relationship between the observed recurrence timescales and the expected stellar rotational periods, the magnetic and variability properties of known magnetic O-type stars, and their implications for understanding wind variability of the broader population.
\end{abstract}

\section{Rotation and large-scale wind variability of O-type stars}

\begin{figure}
\centering
\vcenteredhbox{\includegraphics[width=2.7in]{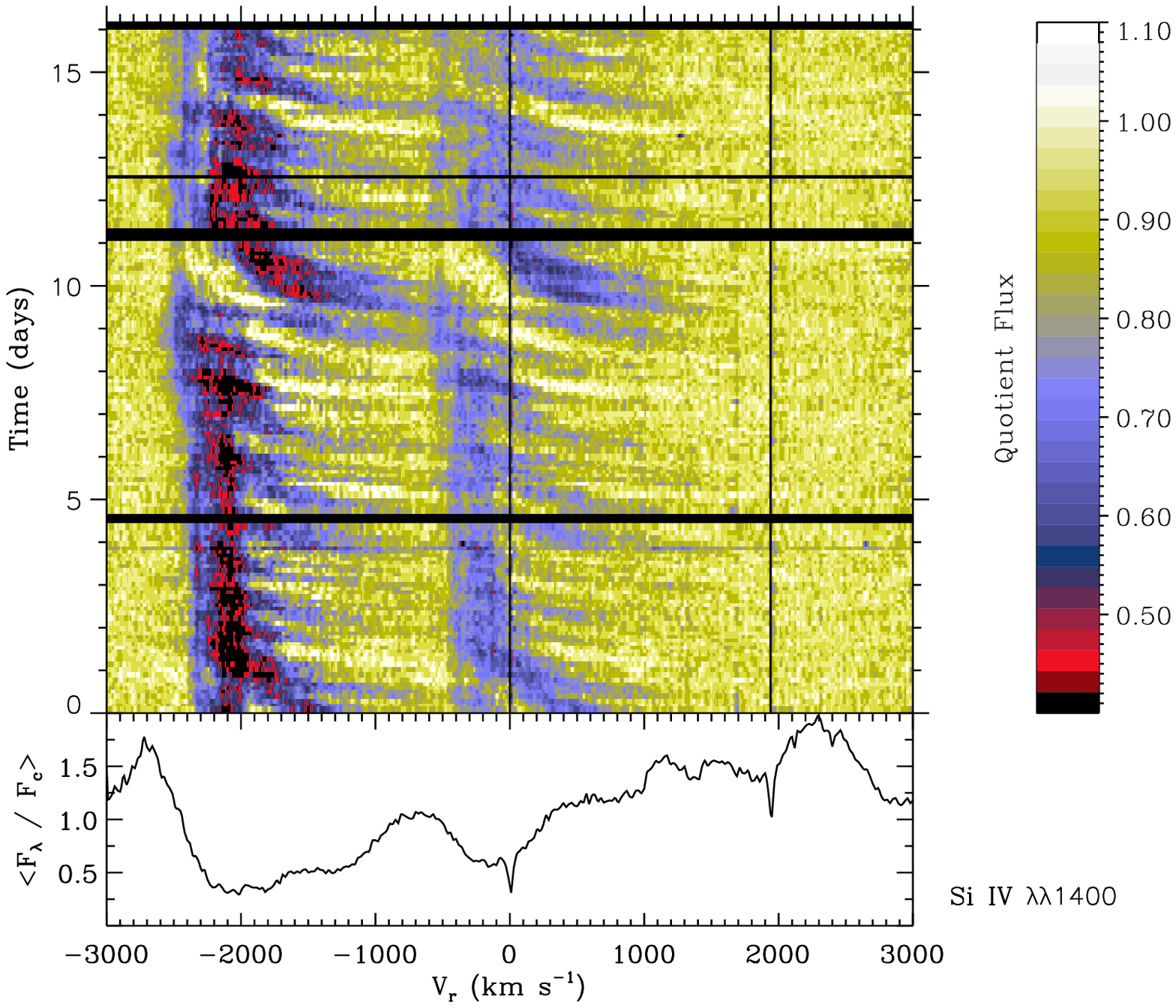}}\hspace{0.5cm}\vcenteredhbox{\includegraphics[width=2.3in]{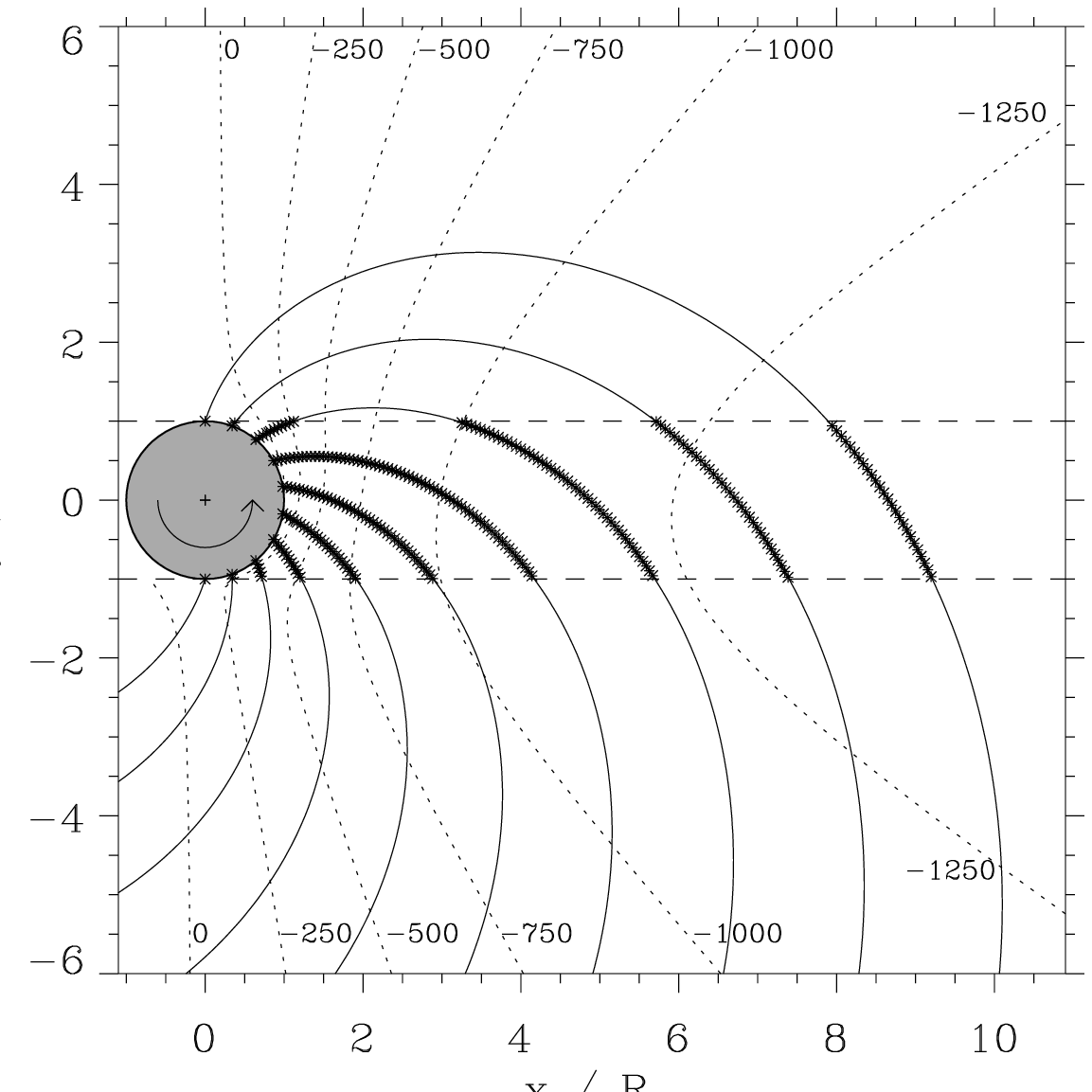}}
\caption{Cyclical variability of O-type stars: the DAC phenomenon. {\em Left -}\ Discrete absorption components in the wind of the O supergiant $\zeta$~Pup \citep{1995ApJ...452L..53M}. \ {\em Right -}\ Schematic illustration of CIRs in the wind of a rotating star \citep{1997A&A...327..699F}. }
\label{dacs}
\end{figure}

Spectroscopic surveys, primarily in the ultraviolet (UV), have demonstrated that the line profiles of essentially all O-type stars exhibit a variety of phenomena (narrow absorption components [NACs], discrete absorption components [DACs], periodic absorption modulations [PAMs], blue edge variability; see \citet{2011IAUS..272..136F}) that have been widely interpreted as diagnostic of large-scale structures in their stellar winds. As the most prominent feature of the wind variability of O-type stars, the DAC phenomenon in particular (see Fig.~\ref{dacs}) has driven a broader discussion of the interpretation of wind-line variations (e.g. \citet{1997A&A...327..281K}, \citet{1997A&A...327..699F}). DACs are narrow, localized optical depth enhancements observed to accelerate slowly (compared to the expectations based on the mean flow of the wind) through the absorption trough of saturated UV absorption lines. A particularly important characteristic of the DAC phenomenon is their recurrence on timescales ($\sim$days) similar to the estimated stellar rotation period (e.g. \citet{1988MNRAS.231P..21P}). Although DACs are always present in multi-epoch observations of the same star and the pattern of variability is always similar, the variations do not seem to maintain a constant phase relationship over intervals of several months. Multi-wavelength observations of a number of OB stars have shown that the cyclical behaviour revealed by DACs is present in various forms in different diagnostics probing the wind down to the surface of the star (e.g. \citet{2001A&A...368..601D},  \citet{1997A&A...327..281K}).

\citet{1995ApJ...453L..37O} and \citet{1996ApJ...462..469C} quantitatively explained this behaviour in terms azimuthally extended structures present in the stellar wind, generated as a consequence of perturbations at the level of the photosphere. These co-rotating interaction regions (or CIRs; first proposed by \citet{1984ApJ...283..303M}) provide a natural explanation for essentially all the properties of cyclical wind variability. 

While the CIR interpretation of wind-line variations is widely accepted, the fundamental outstanding question that remains is that of the origin of the responsible perturbations at the photospheric level. The culprits proposed by \citet{1996ApJ...462..469C}, and those most often discussed in the literature, are non-radial pulsations and magnetic fields.  

Of particular interest in the context of this meeting is the cyclical behaviour of Wolf-Rayet stars. The prototypical example of a cyclically variable WR star is EZ CMa (WR6), a WN4 star which has persistently demonstrated variability with a 3.77~d period. Binarity has been proposed as a potential explanation, although optical, X-ray and radio observations have consistently concluded that CIR-type structures in a perturbed wind provide a better and more coherent explanation of the observations (see e.g. discussion by \citet{1997ApJ...489.1004M}).  \citet{2009ApJ...698.1951S} and \citet{2011ApJ...736..140C} recently published a comprehensive review of WR star variability, examining the optical spectral variability properties of 64 objects. They concluded that large-scale wind variability is a common, but not a ubiquitous, phenomenon amongst WR stars. A significant fraction of their sample display large-scale variations potentially ascribable to CIR-type structures in the stellar wind, while the remainder show small-scale variations or no variability at all.

In the remainder of this paper we explore the idea of magnetic fields as the origin of cyclical wind variability in O-type stars, including WR stars, in light of the most recent observational and theoretical constraints.

\section{Magnetic wind confinement: basic principles}

As reviewed by Petit et al. (these proceedings), the global competition between the outflowing stellar wind (with mass-loss rate $\dot M$ and terminal velocity $v_\infty$) and a large-scale magnetic field (of intensity $B$ at the stellar surface of radius $R$) can be quantitatively characterized by the magnetic wind confinement parameter $\eta_*\equiv {{B^2R^2}\over {\dot M v_\infty}}$. As demonstrated by e.g. \citet{2002ApJ...576..413U}, in regions farther from the star where the wind dominates the interaction, the magnetic field lines are "stretched out" to follow the more-or-less radial wind streamlines. On the other hand, closer to the star where the magnetic field is capable of significantly influencing the wind flow, wind plasma is channeled along closed field lines. For a dipolar magnetic field, this generates a region of characteristic radius $R_{\rm Alf}\simeq \eta_*^{1/4}$, known as the magnetosphere. The characteristics of the wind flow and related phenomena in the magnetosphere are a strong function of the geometry of the magnetic field. For example, in the case of a dipole, wind flows from opposite magnetic hemispheres are expected to be accelerated along closed field lines and to collide at the magnetic equator, generating a strong shock. The hot ($10^{6-8}$~K) plasma produced by this interaction is predicted (and often observed) to emit strongly in both soft and hard X-rays. This is essentially the magnetically-confined wind shock paradigm described by \citet{1997A&A...323..121B}.


As described by \citet{2009MNRAS.392.1022U}, the extended lever arm provided by the magnetic field will also result in enhanced shedding of stellar rotational angular momentum via the wind. In the case of a rotation-aligned magnetic dipole, they find the characteristic spindown timescale $\tau_{\rm spin}={{3}\over{2}}{{kM}\over{BR}}\sqrt{{{v_\infty}\over{\dot M}}}$, where $k$ is the stellar moment of inertia ($\sim 0.1$ for O-type stars). For an hypothetical 30~$M_\odot$ star with a radius of $10~R_\odot$ and wind parameters $\dot M=10^{-7}~M_\odot$/yr and $v_\infty$=2500~km/s, a 1~kG magnetic field corresponds to $\eta_*\simeq 75$ and $\tau_{\rm spin}\simeq 5$~Myr. 

MHD simulations in 2D \citep{2004ApJ...600.1004O} show significantly decreasing global perturbation of the stellar wind with decreasing confinement. Nevertheless, even for $\eta_*$ as small 0.1 the latitudinal dependence of the wind density can be as large as a few. As a consequence, even for sub-critical $\eta_*$, a rotating star with an oblique magnetic field may still show evidence of the presence of the field through (weak) periodic modulation of its wind. (For example, the hypothetical 30~$M_\odot$ star considered above would require a surface magnetic field of about 40~G for $\eta_*\simeq 0.1$.) Therefore, in principle even rather weak magnetic fields are potential candidates for generating the phenomena described in the preceding section. 


\section{Magnetic and variability properties of known magnetic O-type stars}

Only five O-type stars have been confirmed to host magnetic fields: the ZAMS O7V star $\theta^1$~Ori C=HD 37022 (\citet{2002MNRAS.333...55D}, \citet{2006A&A...451..195W}), the more evolved Of?p stars HD 108 (\citet{2010MNRAS.407.1423M}), HD 148937 (\citep{2011A&A...528A.151H}, \citet{2011arXiv1108.4847W}), and HD 191612 (\citet{2006MNRAS.365L...6D}, \citet{2011MNRAS.416.3160W}), and the weak-wind O9V star HD 57682 (\citet{2009MNRAS.400L..94G}, \citet{2011IAUS..272..188G}). In addition, a small number of other O-type stars have been tentatively reported to be magnetic in the modern literature (HD 36879, HD 152408, HD 164794, HD 155806: \citet{2008A&A...490..793H}; HD 37742=$\zeta$~Ori A: \citet{2008MNRAS.389...75B}; HD 155125$=\zeta$~Oph: \citet{2011AN....332..147H}). These stars have either been found through independent observation and/or analysis to be non-magnetic (HD 36879, HD 152408, HD 164794, HD 155806; \citet{2011IAUS..272..182F}, Bagnulo et al. 2011), or have yet to be independently re-observed or re-analysed (HD 37742, HD 155125). These small numbers represent both a reflection of the rarity of O-type stars with detectable magnetic fields, and the challenge of detecting such fields when present.

Three of the 5 known magnetic O stars belong to the Of?p class: early-type O stars exhibiting recurrent, and apparently periodic, spectral variations (in Balmer, He~{\sc i}, C~{\sc iii} and Si~{\sc iii} lines), narrow P Cygni or emission components in the Balmer lines and He~{\sc i} lines, and UV wind lines weaker than those of typical Of supergiants. HD 191612 and HD 108 exhibit strong variability, whereas HD 148937 varies by only weakly.  The variability periods range over several orders of magnitude: whereas the period of HD 148937 is approximately one week (\citet{2008AJ....135.1946N}, \citet{2011arXiv1108.4847W}), the period of HD 191612 is about 1.5 years (\citet{2007MNRAS.381..433H}, \citet{2011MNRAS.416.3160W}), while that of HD 108 is speculated to be of order 50 years \citep{2001A&A...372..195N}. The young star $\theta^1$~Ori C shows variability characteristics and amplitude similar to the Of?p stars, with a period of just over 2 weeks (e.g. \citet{1996A&A...312..539S}, \citet{2008A&A...487..323S}). The spectral variability of the cooler O9V star HD 57682 is somewhat different, exhibiting significant variability of Balmer lines, He~{\sc ii} in absorption and emission, He~{\sc i} in absorption and emission, as well as light elements in absorption (e.g. CNO, Mg, Si), all with a period of about 2 months \citep{2011IAUS..272..188G}.

\begin{figure}
\centering
\vcenteredhbox{\includegraphics[width=1.9in]{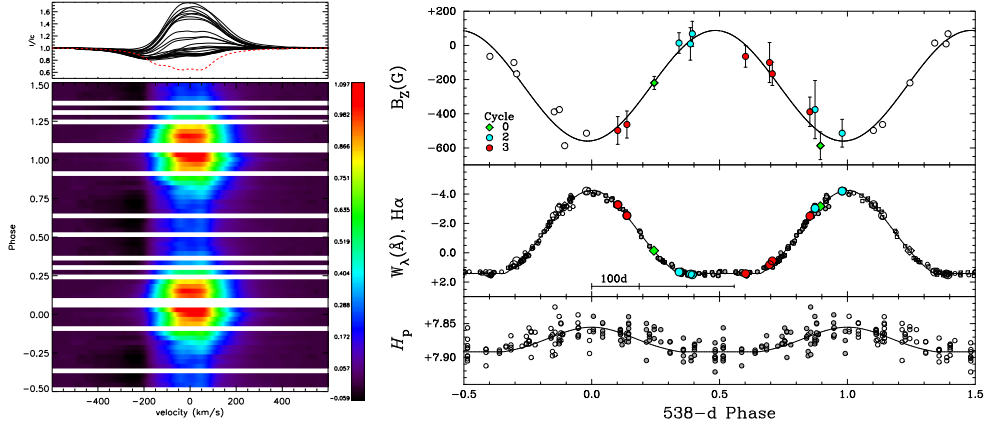}}\hspace{0.0cm}\vcenteredhbox{\includegraphics[width=2.3in,angle=-90]{HD191612_phased.eps}}
\caption{{\em Left -}\ Dynamic spectrum of the H$\alpha$ profile variations of the magnetic Of?p star HD 191612 (e.g. Sundqvist et al. 2011, submitted). {\em Right -}\ Phased variations of the longitudinal magnetic field (upper frame), the H$\alpha$ equivalent width (middle frame) and the $H_{\rm p}$-band photometric brightness (lower frame). The H$\alpha$ equivalent width measurements have been acquired over more than 21 years \citep{2011MNRAS.416.3160W}.}
\label{191612}
\end{figure}

Excluding HD 108\footnote{HD 108 is excluded from this statement because, due to its very long variation period, it is currently not feasible to usefully test the the application of the ORM. However, none of the observations currently in hand suggest any incompatibility with the model.}, each of these stars has been confidently attributed a single variability period that is stable in duration and phase on long timescales. Recent observations and modelling have demonstrated that the variability properties each of these stars can be acceptably understood as magnetic confinement of the stellar wind, within the context of the oblique rotator model (ORM). In this scenario, a large-scale magnetic field (roughly a dipole) is ÓfrozenÓ into the stellar plasma, and tilted relative to the stellar rotation axis. Wind plasma is channelled and structured by the magnetic field above the stellar surface. As the star rotates, observable quantities (e.g. the line-of-sight component of the magnetic field, stellar brightness, absorption and emission lines) are modulated according to the rotational period. 

As illustrated in Fig.~\ref{191612} for HD 191612, the H$\alpha$ variability exhibits a clear phase relationship with the variation of the longitudinal magnetic field. This is interpreted as the result of variable projection of an overdensity of wind plasma, at least partially optically thick, confined by closed field loops in the magnetic equatorial plane (e.g. Sundqvist et al. 2011, submitted). The long-term period and phase stability of the variations implies that the underlying origin of the wind structure - the magnetic field - is similarly stable. 

The physical, wind and magnetic properties of the magnetic O stars are summarized in Table~\ref{ostardata}. Values of $\eta_*$ for these objects range from a few ($\theta^1$~Ori C), to a fews tens (HD 108, 148937, 191612) to more than $10^4$ (HD 57682). Due primarily to uncertainties in the radii and wind parameters of these stars, $\eta_*$ is typically uncertain by roughly one order of magnitude. Spindown times are between 0.5 and 2 Myr except in the case of HD 57682, which is about 10 times longer (due primarily to its low $\dot M$). Spindown timescales are less sensitive to the stellar parameters, and have typical uncertainties of roughly a factor of 3.

In summary, the common properties of all known magnetic O stars are variability with a single, well defined period, consistent phasing of variable quantities over long timescales (years), and slow rotation.

\begin{table}
\centering
\footnotesize
\begin{tabular}{ccccccccccccc}
\hline
\noalign{\smallskip}
ID & ST & $T_{\rm eff}$ & $M_{\rm ev}$ & $R$ & $\log \dot M/\sqrt{f}$ & $v_\infty$ & $P_{\rm rot}$ & $B_{\rm d}$ & $\eta_*$ & $\tau_{\rm spin}$ \\
     &            & (kK)    & ($M_\odot$) & ($R_\odot$) & ($M_\odot$/yr) & (km/s) & (d) & (kG) & &  (Myr)\\ 
\noalign{\smallskip}
\hline
\noalign{\smallskip}
HD 108 & Of?p & 35 & 48.8 & 19.4 & -6.0 & 2000 & O$(10^4)$& $\sim 1$ & 36 & 1.2\\
$\theta^1$~Ori C & O7V & 38 & 23.8 & 5.8 & -6.4 & 2500 & 15.424& 1.3 & 3 &  1.4\\  
HD 57682 & O9V & 35 & 22.0 & 7.0 & -8.9 & 1200 & 63.58& 1.9 & $10^4$&  16.5\\
HD 148937 & Of?p & 40 & 57.9 & 16.6 & -6.0 & 2600 & 7.032& 1.0 & 21 &  1.9\\
HD 191612 & Of?p & 36 & 37.7 &  12.3 & -5.8 & 2700 & 537.6 & 2.5 & 41 &  0.6\\
\noalign{\smallskip}
\hline\hline
\end{tabular}
\label{ostardata}
\caption{Properties of confirmed magnetic O-type stars. Effective temperature $T_{\rm eff}$, evolutionary mass $M_{\rm ev}$, and radius $R$ from Martins et al. (2011). Wind properties, periods and magnetic (dipole) strengths from \citet{2002MNRAS.333...55D}, \citet{2009MNRAS.400L..94G}, \citet{2010MNRAS.407.1423M}, \citet{2011arXiv1108.4847W} and \citet{2011MNRAS.416.3160W}. Calculation of the spindown time assumes $k=0.1$.}
\end{table}


\section{EZ CMa: A case study}

In the context of an exploration of the potential of magnetic wind confinement as the origin of wind modulation of massive stars, the WR star EZ CMa (WR6) represents a particularly interesting case study. This star has been consistently observed to exhibit photometric and spectroscopic variability with a $\sim 3.77$~d period (e.g. \citet{1980ApJ...236L.149M}, \citet{1986AJ.....91..925L}, \citet{1992ApJ...397..277R}, \citet{1994AJ....107.2179A}, \citet{1995ApJ...452L..57S}, \citet{1997ApJ...489.1004M}). However, the detailed nature of the variability (both morphology and phase) differs from season to season, and its origin remains a matter of some debate.

To investigate the hypothesis of a magnetically structured and modulated wind as the origin of the variability of EZ CMa, we can compute the magnetic field strength required to produce various values of the wind confinement parameter $\eta_*$. We assume physical parameters for EZ CMa of $\dot M=10^{-4.6}~M_\odot$/yr, $v_\infty=1700$~km/s and $R=2.8~R_\odot$ (\citet{1999MNRAS.302..499H}, \citet{2004ApJS..154..413M}). For $\eta_*=0.1$, a magnetic field of intensity $B=1.6$~kG is required. For $\eta_*=1$, $B=5$~kG, while for $\eta_*=100$, $B=50$~kG. Due to its enormous mass-loss rate, even relatively small values of $\eta_*$ require large magnetic field strengths. This characteristic is generally true of WR stars.

Recent results constraining the magnetic field of EZ CMa are presented by de la Chevroti\`ere et al. in these proceedings. These authors analyzed high-precision circular polarization spectroscopy of EZ CMa obtained over 4 nights (i.e. approximately the 3.77~d period). Modeling the He~{\sc ii} $\lambda 4686$ line using the model of \citet{2010ApJ...708..615G}, they derived an upper limit of 300~G on the magnetic field at the surface of the hydrostatic core. This upper limit implies a very small value of $\eta_*\ltsim 3.4\times 10^{-3}$. Given the results of MHD simulations, it seems unlikely that such a feeble confinement of the wind is able to generate the rather strong observed modulation. Although the mass of EZ CMa remains uncertain, we can compute the spindown time for de la Chevroti\`ere et al.'s upper limit on the magnetic field using the mass derived spectroscopically by \citet{2006A&A...457.1015H} (19~$M_\odot$). We obtain $\tau_{\rm spin}\gtrsim 2$~Myr, although adopting a mass more consistent with those derived for WR stars in binary systems ($\sim 10~M_\odot$) yields an even shorter $\tau_{\rm spin}\gtrsim 1$~Myr. Therefore, in the scenario in which the 3.77~d period is the stellar rotational period and the modulation is magnetic in nature, the relatively short period may be difficult to reconcile with these calculations and the longer periods of all known magnetic O stars.

\section{Magnetic fields as the origin of DACs}

Observational tests of the proposal that magnetic fields are at the origin of DACs have been carried out by various authors, including \citet{2001A&A...368..601D} ($\xi$~Per), \citet{2008A&A...483..857S} (a sample of 25 O and B stars), and \citet{2009IAUS..259..383H} ($\xi$~Per again). These studies have sought direct evidence of magnetic fields through spectropolarimetric observations, and have achieved longitudinal magnetic field error bars of order 10-100 G. The survey by \citet{2008A&A...483..857S} used the MuSiCoS spectropolarimeter, a less-sensitive predecessor to the current generation of instruments. No magnetic fields were detected. The studies of \citet{2001A&A...368..601D} (also using MuSiCoS) and \citet{2009IAUS..259..383H} (using the newer Narval spectropolarimeter) focused on the O7.5III star $\xi$~Per. For this star, $\eta_*=1$ corresponds to a photospheric field of about 350~G (yielding a peak longitudinal field of about 100~G for a dipolar field), while $\eta_*=0.1$ corresponds to a photospheric field strength of about 100 G (peak longitudinal field of about 30~G for a dipole). The nightly ($1\sigma$) error bars achieved in these studies ($\sim 80$~G and $\sim 30$~G, respectively) demonstrate that the new generation of instrumentation is allowing us to approach the precision needed to usefully test the magnetic hypothesis, at least under restrictive assumptions about the magnetic topology.

\section{Conclusions}

Magnetic fields have been repeatedly proposed as a potential origin for large-scale wind variability observed in O-type stars, including WR stars. Recent MHD simulations demonstrate that a magnetic wind confinement parameter $\eta_*$ as small as 0.1 is capable of producing detectable large-scale periodic modulation of the stellar wind. These same simulations indicate that the "clockwork", phase-locked nature of the modulation - observed in known magnetic O stars over many orders of magnitude of $\eta_*$ - is not expected to disappear for $\eta_*<1$. Therefore, if magnetic confinement is at the origin of wind variability of the broader population of O stars, it seems reasonable to conclude that the character of the magnetic field must be different from the fields of known magnetic O stars. In particular, the lack of phase coherence on long timescales suggests that the field structure may evolve relatively quickly with time. The large-scale nature of the wind variability argues for a similarly large-scale topology of the magnetic field, although there is no {\em a priori} reason to believe that topology must be dipolar. This has implications for the precision required to detect such fields.

Of course, other proposals exist that remain potentially valid explanations of large-scale wind variability. Non-radial pulsations (NRPs) of O stars have timescales generally much shorter than the DAC recurrence timescales \citep{1999LNP...523..305H}, and it has often been claimed for this reason that they are unlikely to be the main cause of DACs. However, \citet{2008ApJ...678..408L} showed that retrograde NRP can lead to recurrence times that are substantially longer than the pulsation period, and even the rotation period of the star. It is therefore possible that the interplay between the pattern speed of multi-mode NRP and the rotation of the star are able to produce variability on the observed timescales, and furthermore prevent DACs from being phase-locked over long intervals (even though the patterns are similar in multi-epoch observations; see, e.g., \citet{1999A&A...344..231K}). And evidence exists that some DAC stars exhibit non-radial pulsations (e.g. \citet{1999A&A...345..172D}), as do some WR stars (e.g. \citet{2011ApJ...735...13H}).

\section*{Acknowledgements} 

Thanks to V. Petit, A. ud-Doula, A. Fullerton J. Grunhut and A. Moffat for comments and feedback that contributed to the improvement of this manuscript.

\bibliography{wade}

\end{document}